\documentclass[aip,jcp,reprint,numerical,english,floatfix]{revtex4-1}
\usepackage{xcolor}
\usepackage{pdfcolmk}
\usepackage{amsmath}
\usepackage{amssymb}
\usepackage{graphicx}
\usepackage{esint}
\usepackage{ulem}

\usepackage{array}
\usepackage{bm}
\usepackage{amsmath}
\usepackage{graphicx}
\usepackage{babel}

\usepackage{color}
\usepackage{array}
\usepackage{textcomp}
\usepackage{multirow}

\makeatletter

\providecommand{\tabularnewline}{\\}

\@ifundefined{textcolor}{}
{%
 \definecolor{BLACK}{gray}{0}
 \definecolor{WHITE}{gray}{1}
 \definecolor{RED}{rgb}{1,0,0}
 \definecolor{GREEN}{rgb}{0,1,0}
 \definecolor{BLUE}{rgb}{0,0,1}
 \definecolor{CYAN}{cmyk}{1,0,0,0}
 \definecolor{MAGENTA}{cmyk}{0,1,0,0}
 \definecolor{YELLOW}{cmyk}{0,0,1,0}
}

\makeatother

\begin{document}
\global\long\def\icm{\mathrm{cm}^{-1}}
\global\long\def\Qy{{\rm Q}_{y}}
\global\long\def\Qx{{\rm Q}_{x}}

\title{Coherence and population dynamics of chlorophyll excitations in FCP
complex: Two-dimensional spectroscopy study}

\author{Vytautas Butkus}

\affiliation{Department of Theoretical Physics, Faculty of Physics, Vilnius University,
Sauletekio ave. 9-III, 10222 Vilnius, Lithuania}

\affiliation{Center for Physical Sciences and Technology, Savanoriu ave. 231,
02300 Vilnius, Lithuania}

\author{Andrius Gelzinis}

\affiliation{Department of Theoretical Physics, Faculty of Physics, Vilnius University,
Sauletekio ave. 9-III, 10222 Vilnius, Lithuania}

\affiliation{Center for Physical Sciences and Technology, Savanoriu ave. 231,
02300 Vilnius, Lithuania}

\author{Ram\={u}nas Augulis}

\affiliation{Center for Physical Sciences and Technology, Savanoriu ave. 231,
02300 Vilnius, Lithuania}

\author{Andrew Gall}

\affiliation{Institut de Biologie et Technologies de Saclay, Bât 532, Commissariat
à l'Energie Atomique Saclay, 91191 Gif sur Yvette, France}

\author{Claudia B\"{u}chel}

\affiliation{Institut für Molekulare Biowissenschaften, Universität Frankfurt,
Max-von-Laue-Straße 9, Frankfurt, Germany}

\author{Bruno Robert}

\affiliation{Institut de Biologie et Technologies de Saclay, Bât 532, Commissariat
à l'Energie Atomique Saclay, 91191 Gif sur Yvette, France}

\author{Donatas Zigmantas}

\affiliation{Department of Chemical Physics, Lund University, P.O. Box 124, 22100
Lund, Sweden}

\author{Leonas Valkunas}

\affiliation{Department of Theoretical Physics, Faculty of Physics, Vilnius University,
Sauletekio ave. 9-III, 10222 Vilnius, Lithuania}

\affiliation{Center for Physical Sciences and Technology, Savanoriu ave. 231,
02300 Vilnius, Lithuania}

\author{Darius Abramavicius}

\email{darius.abramavicius@ff.vu.lt}

\affiliation{Department of Theoretical Physics, Faculty of Physics, Vilnius University,
Sauletekio ave. 9-III, 10222 Vilnius, Lithuania}
\begin{abstract}
The energy transfer processes and coherent phenomena in the fucoxanthin--chlorophyll
protein complex, which is responsible for the light harvesting function
in marine algae diatoms, were investigated at 77 K by using two-dimensional
electronic spectroscopy. Experiments performed on the femtosecond
and picosecond timescales led to separation of spectral dynamics,
witnessing evolutions of coherence and population states of the system
in the spectral region of $\Qy$ transitions of chlorophylls $a$
and $c$. Analysis of the coherence dynamics allowed us to identify
chlorophyll (Chl) $a$ and fucoxanthin intramolecular vibrations dominating
over the first few picoseconds. Closer inspection of the spectral
region of the $\Qy$ transition of Chl~$c$ revealed previously not
identified mutually non-interacting chlorophyll $c$ states participating
in femtosecond or picosecond energy transfer to the Chl~$a$ molecules.
Consideration of separated coherent and incoherent dynamics allowed
us to hypothesize the vibrations-assisted coherent energy transfer
between Chl~$c$ and Chl~$a$ and the overall spatial arrangement
of chlorophyll molecules.
\end{abstract}


\maketitle

\section{Introduction}

Diatoms are unicellular chromophyte algae inhabiting marine environment.
They are the major players in photosynthesis, accounting for nearly
a quarter of the global primary production\citep{Falkowski_science_1998,Field_science_1998,Mann1999}.
Light-harvesting in diatoms is performed by their intrinsic membrane
proteins, such as the fucoxanthin--chlorophyll protein (FCP) complex.
FCP shares some structural and sequence homology with the light-harvesting
complex II (LHCII) from the higher plants\citep{FCP_homology_1998},
but contains more carotenoids per chlorophyll. Carotenoids fucoxanthins
(Fx) strongly absorb blue--green photons with the longest propagation
distances in water. Such adaptation allows diatoms to efficiently
harvest sunlight in underwater conditions.

Information on the molecular structure of FCP is very limited; there
is no crystallographic data of the complex. Several preliminary models
of pigment organization in FCP, based on the assumption that the structure
of FCP is similar to the LHCII, have been proposed\citep{Papagiannakis2005,Wilhelm_FCP_model_protist_2006,FCP_RR_bba_2010,Gildenhoff2010a}.
These models, however, are currently limited for several reasons.
First, only some pigment binding sites are conserved between FCP and
LHCII. Second, the pigment composition and, thus, detailed functional
roles in LHCII and FCP are markedly different, as the latter contain
nearly as many fucoxanthin molecules as chlorophylls (Chl)---most
probably 8 Fx, 8 Chl~$a$ and 2 Chl~$c$ molecules\citep{Premvardhan2009,FCP_RR_bba_2010}---while
LHCII is composed of 8 Chl~$a$, 6 Chl~$b$ and 4 xanthophyll molecules\citep{Liu-Chang2004}. 

Chl~$c$ molecules enhance the absorption of blue photons in FCP,
as well\citep{Croce2014}. Ultrafast molecular excitation transfer
from Chl~c to Chl~$a$ molecules was shown to occur at room temperature\citep{Songaila_FCP_2013}.
Nonetheless, the whole picture of the energy funnels in the FCP complex
and the physical phenomena ensuring ultrafast energy transfer still
needs to be detailed. Unfortunately, lack of structural information
about FCP prevents detailed numerical studies. All the information
about the inner workings of this complex comes solely from the experimental
data and its analysis, which is often done in the light of the proposed
models\citep{Papagiannakis2005,Wilhelm_FCP_model_protist_2006,FCP_RR_bba_2010,Gildenhoff2010a}.
Therefore, it is important to obtain as much experimental data as
possible to constrain and refine the current models.

The two dimensional (2D) electronic spectroscopy (ES) is a remarkable
tool for the analysis of coherent and incoherent phenomena appearing
in molecular excitation dynamics. The first and the most recognized
results of 2D ES demonstrating complex behavior of the excitation
evolution were delivered for the photosynthetic Fenna--Matthews--Olson
(FMO) complex from purple bacteria\citep{Brixner-fleming-nature2005}.
Since then, 2D ES has been successfully applied for studies of a number
of different systems; 2D ES essentially helped to establish or update
existing kinetic schemes of light-harvesting antennae and reaction
centers of plants\citep{Schlau-Cohen_LHCII_2009,Gelzinis_PS2RC_2D_tight_binding,Fuller_NChem_2014,Romero2014}
and bacteria\citep{Lee-Fleming2007,ZigmantasFlemingPNAS2006,Dostal2012a}.
The 2D ES was also performed on the whole photosystem I complex demonstrating
signatures of fast energy transfer\citep{Anna_Scholes_PSI_2D_jpcl_2012}.
In addition, it was utilized in studies of other molecular aggregates,
for instance, polymers\citep{Collini2009}, cylindrical (bi-tubular)
J-aggregates\citep{Milota200945}, quantum dots\citep{Turner2012}
etc.

Most of these findings were obtained from experiments and theoretical
simulations performed using (or assuming) the knowledge of the structural
organization of the pigment molecules (often with resolution of a
few angströms). Thus, these studies were valuable not only for widening
a general understanding of the physical mechanisms within molecular
systems, but also were convenient for testing and interpreting the
outcomes of the new method of 2D ES. As the method itself has already
been well approved and appreciated, it becomes useful to apply the
developed tools of analysis of 2D spectra for more obscured molecular
systems, for instance the FCP complex, and try to obtain more information
about their structure/composition. 

Recently, a few multidimensional spectroscopy studies of FCP reported
valuable insight about various unknown characteristics of the complex.
A novel method of two-color two-dimensional spectroscopy allowed to
map the the carotenoid--chlorophyll energy transfer pathways and resolved
spectral heterogeneity of carotenoids in FCP\citep{Gelzinis_FCP_2D2color_BBA_2015}.
It was found, that fucoxanthins in FCP are spectrally distinguishable
and yet transfer energy to chlorophylls very efficiently. In our recent
study\citep{Songaila_FCP_2013}, dynamics of two-dimensional spectra
of Chl~$a$ and Chl~$c$ ${\rm Q}_{y}$ absorption bands at room
temperature were analyzed and we reported ultrafast energy transfer
between Chl~$c$ and Chl~$a$, with signatures of excitonic coupling.
However, limited temporal and spectral resolution in these particular
experiments did not provide the possibilities to fully resolve coherent
dynamics or a more fine structure of the FCP spectrum. 

Here we present results of the 2D ES of FCP obtained with higher spectral
and temporal resolution than in our previous work\citep{Songaila_FCP_2013}.
To better discriminate the weak spectral features of FCP, experiments
were carried out at $77$~K instead of the room temperature. The
2D spectra were analyzed with both femtosecond and picosecond time
steps, thus, providing us the possibility to separate coherent and
incoherent processes. In this work, we show that there are at least
two Chl~$c$ molecules in FCP, which are clearly spectrally resolvable
and participate in energy transfer on remarkably different timescales.
We discuss that this might be related to certain spatial arrangement
of Chl~$a$ and $c$ molecules in the complex. Finally, the analysis
of rich coherent dynamics during the first picosecond reveals signatures
of Chl~$a$ and fucoxanthin intramolecular vibrations, the former
of which might play a role in enhancing ultrafast Chl~$c$-to-$a$
energy transfer.

\section{Materials and methods}

Fucoxanthin--protein complex was extracted and purified from diatom
\emph{Cyclotella meneghiniana}. The procedure of purification and
extraction is given in detail in Ref.~\citep{Buchel2003}. The sample
for the measurements was diluted with the glycero--buffer--detergent
solution (10~mM Mes pH~6.5 buffer, 2~mM KCl, 0.03\% (w/v) n-Dodecyl
$\beta$-D-maltoside and 60\% (v/v) glycerol) to achieve optical density
of about 0.3 at Chl~$a$ $\Qy$ band in 0.5~mm cell.

The sample was placed in a 0.5~mm quartz cell and gradually cooled
to 77~K in a liquid ${\rm N_{2}}$ bath cryostat (Oxford Instruments).

The setup for coherent two-dimensional electronic spectroscopy was
based on the inherently phase-stabilized design by Brixner et al.\citep{Brixnerstiopkin-fleming-optlett2004}.
In more detail, 10~fs, 100~nm full width at half maximum (FWHM)
light pulses (3~nJ energy per pulse at the sample) centered at 615~nm
were generated by home built non-collinear optical parametric amplifier
were pumped by the Yb:KGW amplified laser system (“Pharos”, Light
Conversion Ltd.) with repetition rate reduced to 7~kHz by means of
pulse picker. Diffractive--optics--based non-collinear four-wave mixing
setup with phase-matched box geometry was used. The third-order signal
was obtained by spectral interferometry with the help of a heterodyning
local-oscillator pulse. Sensitivity and noise resistance was significantly
enhanced by means of double modulation lock-in detection\citep{Augulis_Zigmantas_2011}.
Coherence time delay was scanned within -140--300~fs time range with
2~fs time step controlled by movable fused-silica wedges in beams
$\vec{k}_{1}$ and $\vec{k}_{2}$. 2D spectra were obtained by processing
the accumulated sequence by the Fourier transform method\citep{Hybl1998}.
The absolute phase of the 2D spectrum was retrieved by matching its
projection with the transient absorption spectrum\citep{Jonas_2003}
recorded with the same setup with only two beams (for pump and probe)
in use. Real (absorptive) parts of the total (rephasing+nonrephasing)
spectra were used in further analysis.

\section{Results}

\begin{figure}
\includegraphics{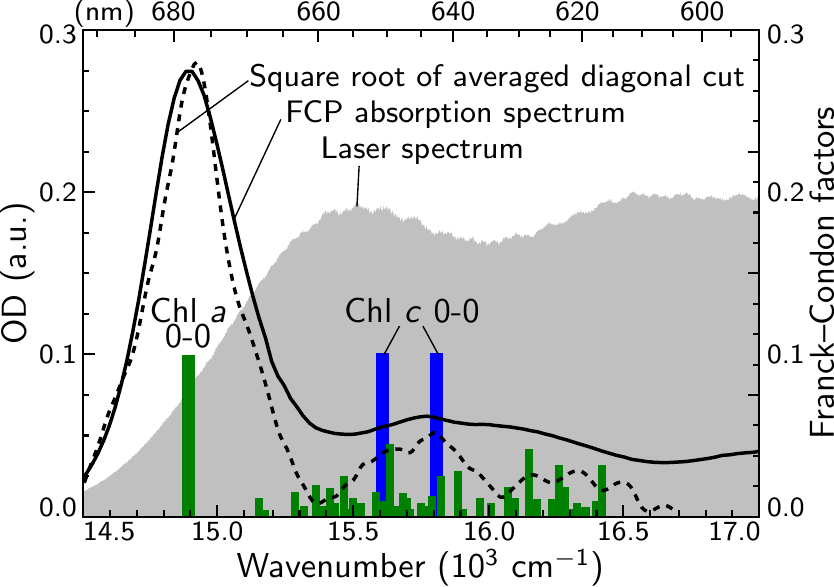}

\protect\caption{\label{fig:Absorption-spectrum-of}The absorption spectrum of FCP
at 77~K (solid black line), square root of averaged (delay times
from $t_{2}=30$~fs to 800~fs) diagonal cut of the 2D spectrum (dashed
black line), excitation laser spectrum of 2D~ES experiment (filled
gray area). Thicker vertical green and blue sticks indicate the energies
of the 0--0 $\protect\Qy$ transitions of Chl~$a$ and Chl~$c$,
respectively. Vibrational states of Chl~$a$, proposed by the hole
burning measurements\citep{Gillie1989} are shown by thinner green
vertical sticks. Their heights indicate the Franck--Condon factors
of the corresponding vibrations (values on the right axis).}
\end{figure}

The absorption spectrum of FCP measured at 77~K is shown in Fig.~\ref{fig:Absorption-spectrum-of}
by the black solid line. The spectrum contains a well expressed band
at $14895\,\icm$ and a few weaker features around $15700\,\icm$
and $17000\,\icm$. The dominating band at $14895\,\icm$ is known
to represent the $\Qy$ transition of the core pigment Chl~$a$.
The feature at $15680\,\icm$ was assigned to the $\Qy$ transition
of Chl~$c$, while the origin of bands at around $16200\,\icm$ and
$17000\,\icm$ were discussed to be reflecting the $\Qx$ transitions
of Chl~$a$ and Chl~$c$, respectively\citep{Songaila_FCP_2013}.

The spectrum of the laser pulse used in the 2D ES experiment is shown
by a gray shaded area in Fig.~\ref{fig:Absorption-spectrum-of}.
The spectrum was centered at around 16240~$\icm$ with the FWHM of
$\sim2600\,\icm$. The laser spectrum was intentionally shifted to
the blue to enhance the signals form the weak absorption bands of
FCP: while it only partly covered the main absorption band at $14895\,\icm$,
the other weak features were fully covered by the laser spectrum.

Two different 2D ES experiments were carried out. In the first one,
the 2D spectra at delay times $t_{2}$ up to 800~fs were taken at
every 10~fs. Such a high temporal resolution is very useful for the
analysis of the ultrafast excitation transfer dynamics as well as
for studies of the quantum coherences and allowed us to resolve coherent
beatings up to $1668\,\icm$ in the frequency domain. The second set
of data consisted of the 2D spectra measured at delay times $t_{2}$
up to 1~ns with different time steps: from 50~fs for the initial
evolution to 100~ps for delay times longer than 100~ps. These measurements
were used for the analysis of the incoherent long-time population
dynamics in FCP.

\begin{figure*}
\includegraphics{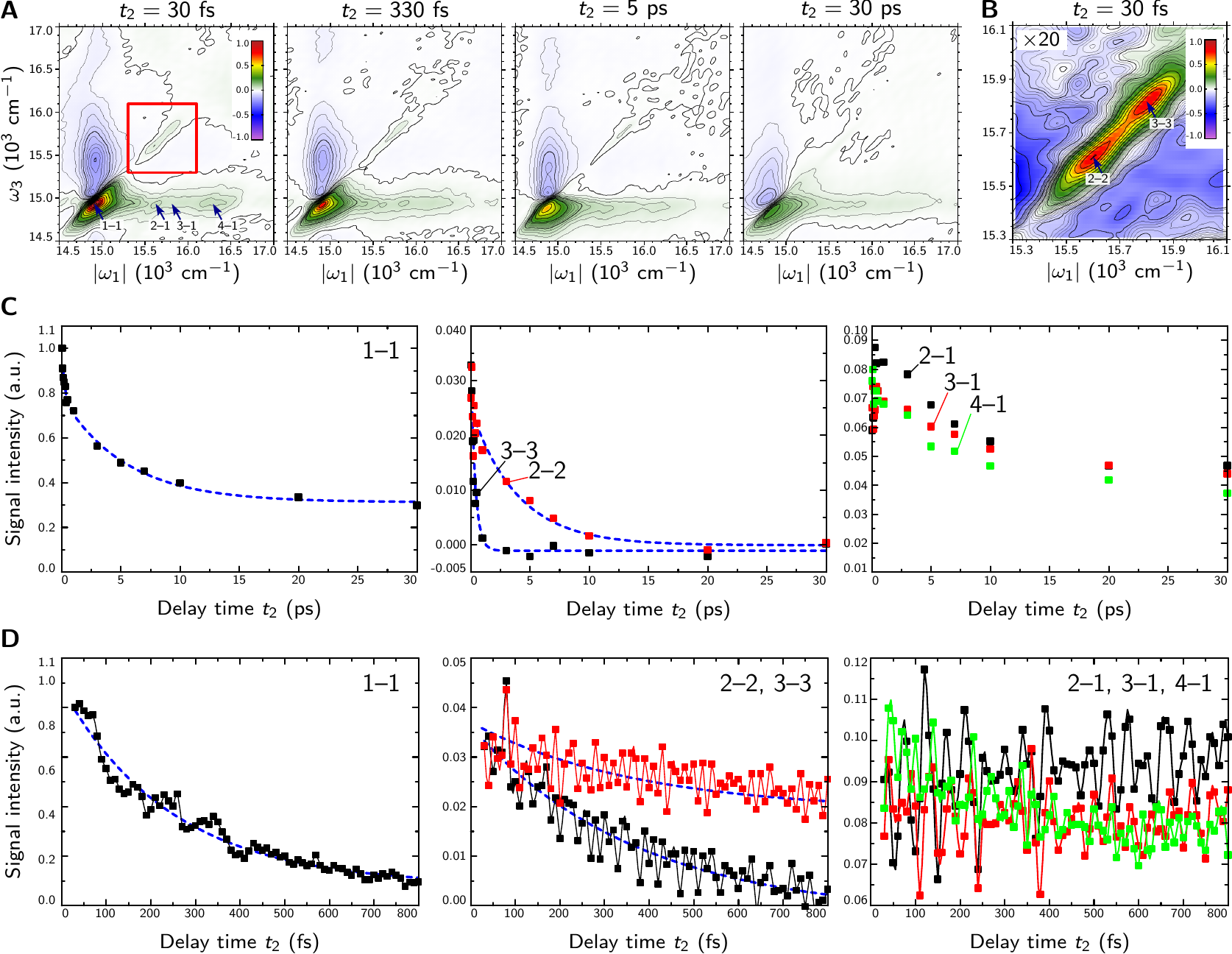}

\protect\caption{\label{fig:2D-absorptive-spectra}\label{fig:twopeak}(A) 2D absorptive
spectra at different delay times. All the spectra are normalized to
the maximum of the spectrum at $t_{2}=30$~fs and drawn using the
same scale. Thicker contour lines are drawn at every $10\%$ of the
maximum of the scale; thinner lines are drawn every $2\%$, up to
$\pm20\%$. (B) The two-peak structure in the 2D spectrum of FCP.
Zoomed-in region, as shown in Fig.~\ref{fig:2D-absorptive-spectra}A
by the red square. Values in the spectrum are multiplied by 20 and
the spectrum is drawn using the same scale as in (A). (C) Picosecond
and (D) femtosecond time traces of a few selected peaks indicated
by arrows in (A) and (B). Values obtained by averaging around the
selected points within the circle of $100\,\protect\icm$ radius are
presented. Multi-exponential fits of dynamics of a few peaks are shown
by the blue dashed lines (see the text for the fitting parameters).}
\end{figure*}

A few 2D spectra at different delay times ($t_{2}=30$~fs, $330$~fs,
$5$~ps and $30$~ps) are depicted in Fig.~\ref{fig:2D-absorptive-spectra}A.
In the spectra, the main band corresponding to the Chl~$a$ transition
at $\omega_{3}=\omega_{1}=14926\,\icm$ dominates. The fact that the
band is not observed exactly at $14895\,\icm$ as in the absorption
spectrum, but rather blue-shifted by $31\,\icm$ is caused by the
excitation laser spectrum being slightly off-resonant with respect
to the main FCP absorption band (see Fig.~\ref{fig:Absorption-spectrum-of}).
The negative excited state absorption is well resolved above the diagonal
in the 2D spectrum in the vicinity of the main positive absorption
peak. The excited state absorption features in the 2DES could arise
both due to excitonically coupled pigments (chlorophylls) and the
corresponding double-excitation transitions\citep{butkusJagg,Abramavicius_epl_Jagg_2013,Han_Zhang_Abramavicius_Jagg_2D_jcp_2013},
as well as from the double-excitations of chlorophylls themselves\citep{Du2011}.

In the 2D spectra shown in Fig.~\ref{fig:2D-absorptive-spectra},
thinner contour lines, drawn at every 2\% of the maximum highlight
the fine structure of the diagonal and off-diagonal peaks (cross-peaks)
away from the main Chl $a$ absorption band. The most pronounced features
correspond to the cross-peaks lying around the same emission frequency
$\omega_{3}$ as the Chl~$a$ absorption band. These peaks might
either signify the excitonic coupling between Chls $a$ and $c$\citep{Songaila_FCP_2013},
or they might be caused by the Chl~$a$ vibrational modes.

In the region of the $\Qy$ of the Chl $c$ transition frequency,
i.e. at around $\omega_{3}=\omega_{1}\approx15700\,\icm$, two diagonal
peaks are clearly resolved at excitation/emission frequencies $15610\,\icm$
and $15810\,\icm$. The positions of these peaks are indicated by
two vertical bars in the absorption spectrum shown in Fig.~\ref{fig:Absorption-spectrum-of},
coinciding with the maxima of the diagonal cut of the averaged 2D
spectrum (black dashed line in Fig.~\ref{fig:Absorption-spectrum-of}).
The zoomed-in region of the 2D spectrum from $\omega_{3}=\omega_{1}=15300\,\icm$
to $\omega_{3}=\omega_{1}=16100\,\icm$ (shown by a red square in
Fig.~\ref{fig:2D-absorptive-spectra}A) is depicted in Fig.~\ref{fig:twopeak}B
at delay time $t_{2}=30$~fs. No cross-peak associated with these
two transitions is visible at any delay time disproving the presence
of excitonic coupling between states at $15610\,\icm$ and $15810\,\icm$.

The 2D spectra contain rich dynamics both at short (femtosecond) and
long (picosecond) delay times, as it was extracted from the separate
measurements described above. Dynamics of a few selected peaks in
the spectra, occurring during first 30~ps are shown in Fig.~\ref{fig:2D-absorptive-spectra}C.
The main diagonal peak (indicated by ``1--1'') decays with timescales
of $220\pm100$~fs and $5.4\pm0.6$~ps. The cross-peaks (indicated
by ``2--1'', ``3--1'' and ``4--1'') decay with the same longer
rate of $5.4$~ps as the ``1--1'' peak; the initial evolution is
dominated by the high amplitude coherences (Fig.~\ref{fig:2D-absorptive-spectra}D)
and, thus, shorter lifetimes could not be unambiguously evaluated.
Diagonal peaks at $15610\,\icm$ and $15810\,\icm$ show completely
different long-time dynamics. The upper peak decays much slower than
the lower one, as it can be seen by the spectral traces in Fig.~\ref{fig:twopeak}C.
We could fit the dynamics of both peaks as single-exponential decays
with timescales of $\tau_{15610}=320\pm70$~fs and $\tau_{15810}=3.9\pm1.6$~ps,
respectively. For longer delay times, additional exponential decays
are present.

The short-time (femtosecond) evolution of the same points in the spectrum
are shown in Fig.~\ref{fig:2D-absorptive-spectra}D demonstrating
the oscillating behavior of the the diagonal and off-diagonal peaks
corresponding to the Chl~$a$. Coherence beatings can be analyzed
by performing the Fourier transform of the time traces of the peaks.
A very convenient way to represent the overall complexity of oscillations
in the 2D spectrum is through the Fourier maps\citep{Panitchayangkoon2011,Cundiff-3Dspec-ncomms2012,Butkus_cpl_2012,Seibt-JPCC2013,Novoderezhkin_maps2014,Camargo2015}.
They are constructed by performing the Fourier transform of the each
data point in the 2D spectrum over the delay time $t_{2}$ after subtraction
of slow exponential decay contributions. Then, the Fourier amplitude
can be plotted as a two-dimensional map against excitation and detection
frequencies $\omega_{1}$ and $\omega_{3}$ for a fixed Fourier frequency
$\omega_{2}$. Each Fourier map then can be integrated over $\omega_{1}$
and $\omega_{3}$ and the integral value represented as function of
coherence frequency $\omega_{2}$. Such a dependency would show the
averaged distribution of various oscillation frequencies. 

\begin{figure}
\includegraphics{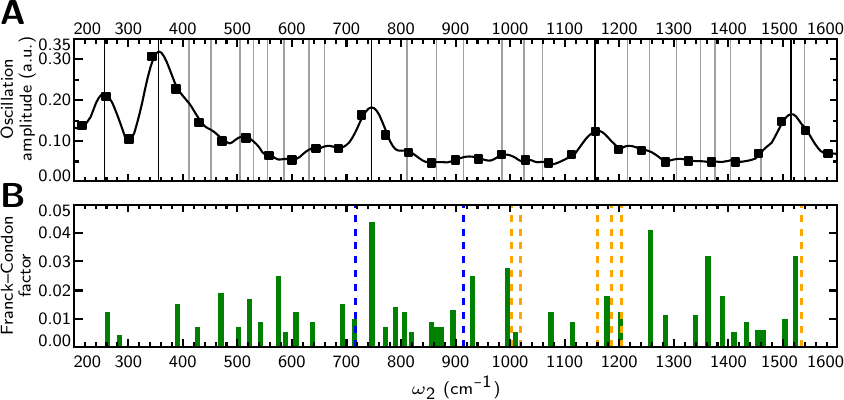}

\protect\caption{\label{fig:frobenius}(A) Amplitude of oscillations (Fourier map norm)
as a function of coherence frequency $\omega_{2}$. Dependency extracted
from Fourier transforms of zero-padded and not-padded spectral time
evolutions are shown by symbols and lines, respectively. Determined
frequencies of oscillations are shown by the vertical black and gray
lines. (B) Franck--Condon factors of vibrational intramolecular modes
of Chl~$a$ from the hole burning experiment\citep{Gillie1989} (green
sticks), proposed frequencies of electronic coherences (blue dashed
lines) and vibrational Raman frequencies of fucoxanthin (orange dashed
lines).}
\end{figure}

We chose to represent the amplitude of observed oscillations by the
value of the Fourier map norm, which is equal to the square root of
the sum of squares of the data points of Fourier maps at frequency
$\omega_{2}$. From the oscillation amplitude dependency shown in
Fig.~\ref{fig:frobenius}A, 5 sharp peaks at $\omega_{2}=255\,\icm,$
$345\,\icm$, $745\,\icm$ $1155\,\icm$ and $1515\,\icm$ can be
readily pointed out. More rigorous analysis of the second derivative
of the oscillation amplitude function reveals 24 more frequencies.

\section{Discussion}

The 2D ES spectra obtained with femtosecond and picosecond time resolutions
give a wealth of information regarding both evolution of population
and coherent states of the molecular system under consideration. Special
attention should be taken into consideration when analyzing the processes
occurring on different timescales. Thus, before turning to the analysis
of the femtosecond evolution of the measured spectra, let us discuss
a few mechanisms that contribute to the spectral lineshape formation:
pulse overlap, spectral diffusion, dynamic Stokes shift and coherence
beatings. They can be important factors that might lead to ambiguities
and errors when interpreting the time-dependencies of the 2D spectrum.

Pulse overlap effects dominate during the initial delay time evolution,
comparable to the length of the laser pulses. That is, besides the
double-coherence pathways, additional rephasing contributions appear
in the phase matching direction when the $\vec{k}_{1}$ and $\vec{k}_{2}$
pulses do overlap\citep{Abramavicius_dimer_2010}. Therefore, the
initial evolution of the 2D spectrum (in our case first 30~fs) is
dominated by the sharp decay of these additional pulse-overlap contributions
and should be considered very carefully as they could be confused
with some ultra-fast relaxation mechanisms.

Spectral diffusion can also make an evident influence on the 2D spectrum\citep{Abramavicius-EPL2007}.
It accounts for the loss of correlation of the narrow-bandwidth excitation
and spectrum is manifested by the increase of the anti-diagonal width
of the peaks in the absorptive 2D spectrum with the delay time. The
spectral diffusion affects both the rephasing and nonrephasing spectral
contributions. For a simple two-level system the rate of the spectral
diffusion can be related to the bath relaxation time and results in
exponential decrease of the height of the main absorption peak in
the 2D spectrum\citep{hamm2011concepts}. As such an exponential evolution
overlaps with the other temporal lineshape formation mechanisms, it
becomes troublesome not to overinterpret the spectral diffusion as
a relaxation-type signal. Comparing the evolution of the peak height
against its anti-diagonal width provides a possible way to evaluate
the contribution of the spectral diffusion. However, sometimies this
is not possible because of overlaps of closely-positioned coupled
states and cross-peaks. The other method is to check the time evolution
of the integral over some area of the spectrum in the vicinity of
the peak under consideration in the absorptive spectrum. If the radius
of the integrated area is larger than the homogeneous linewidth, the
spectral diffusion is averaged out from such signal.

System states during the delay time $t_{2}$ experience relaxation
processes through the vibronic pathways. Such a vibrational cooling
is often referred to as the dynamic Stokes shift and expressing itself
as the shifting of peaks down in the energy scale along the emission
axis with $t_{2}$ in the 2D spectrum. As the ground state bleaching
and excited state emission contributions can, in principle, experience
the dynamic Stokes shift on different timescales, it can cause some
overlaps of these positive and negative contributions, giving a complex
time-dependence of both diagonal peaks and cross-peaks. Moreover,
the timescale of the dynamic Stokes shift can be around a few hundreds
of femtoseconds -- therefore, without a careful analysis such evolution
might once again be wrongly assigned to some ultrafast relaxation
pathway.

Finally, the initial evolution of the 2D spectrum is seriously affected
by the coherence beatings. Lifetimes of the electronic coherences
can be as short as tens of femtoseconds and vibrational/vibronic coherences
can result in oscillations for as long as a few picoseconds. Although
static disorder tends to dephase electronic coherences very quickly,
their contribution to the initial dynamics can still be substantial.
For example, in spectra recorded with a low $t_{2}$ resolution, the
coherent signal adding on top of the population evolution can be as
significant as the population signal itself. Thus, coherence beatings
can seriously impede interpretations of the population dynamics if
the 2D spectrum is of low resolution in delay time $t_{2}$.

\subsection{Identification of coherences}

In the 2D ES spectra of FCP we would expect coherences of various
origins and amplitudes to appear: ground state vibrational coherences
of Chl~$a$, Chl~$c$ and Fx, excited state vibronic coherences
of the same frequencies, mixed coherences of shifted or combinational
frequencies\citep{Fuller_NChem_2014,Butkus-nanoring2015} and electronic
coherences due to excitonic coupling of chlorophylls. Such distinction
between the coherences of different origin is conveniently observed
in the Fourier amplitude maps. In the rephasing spectrum, ground state
coherences are manifested by strong beatings below and on the diagonal\citep{Butkus_cpl_2012}
while the corresponding vibronic coherences -- symmetrically above
the diagonal\citep{Novoderzhkin_PS2RC2005,Butkus2013,Butkus_vibrations_theory_jcp_2014}.
Mixed coherences, that represent superpositions of quantum mechanically
mixed states, are observed as complex and asymmetric patterns of oscillating
peaks in the Fourier maps\citep{Butkus_vibrations_theory_jcp_2014,Butkus-nanoring2015}.
Electronic coherences in the rephasing spectrum manifest themselves
exclusively as oscillatory cross-peaks and no oscillations on the
diagonal and vice versa for the nonrephasing spectrum\citep{Butkus_cpl_2012}.
However, oscillations in the nonrephasing contribution of the spectrum
are more sensitive to static disorder of pigment energies and, thus,
in the systems with large inhomogeneous broadening coherences from
the rephasing contributions will dominate over those from the nonrephasing
contribution.

Let us discuss about the origin of coherences in the spectra of FCP
by comparing the obtained dependencies of the intensity of oscillations
(Fig.~\ref{fig:frobenius}A) with known frequencies corresponding
to the intramolecular vibrational modes of Chl~$a$ . We refer to
the hole burning experiments on single chlorophylls carried out in
1.6~K by Gillie~et~al.\citep{Gillie1989} (green sticks in Fig.~\ref{fig:frobenius}B)
and a recent study of Du~et~al.\citep{Du2011}, in which the frequencies
were obtained from the time and frequency resolved pump--probe spectra
at 297~K by the Fourier analysis, similar to the one employed here.

\begin{figure*}
\includegraphics{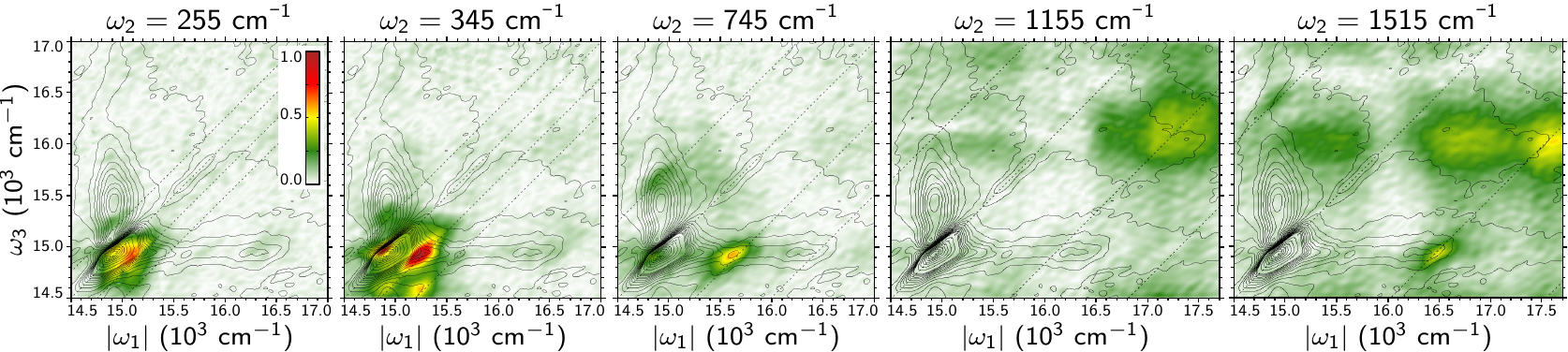}\protect\caption{\label{fig:MAPS}Fourier amplitude maps at coherence frequencies $\omega_{2}=255\,\protect\icm$,
$345\,\protect\icm$, $745\,\protect\icm$, $1155\,\protect\icm$
and $1515\,\protect\icm.$ Fourier amplitude is shown by a color scale,
normalized to the maximum of the map at $\omega_{2}=345\,\protect\icm$.
Black contour lines show the 2D spectrum at $t_{2}=30$~fs. Diagonal
and parallel to the diagonal dashed lines are shifted by the corresponding
value of $\omega_{2}$. Notice that excitation axex of maps at $\omega_{2}=1155\,\protect\icm$
and $1515\,\protect\icm$ are extended compared to the other maps.}
\end{figure*}

\begin{figure}
\includegraphics{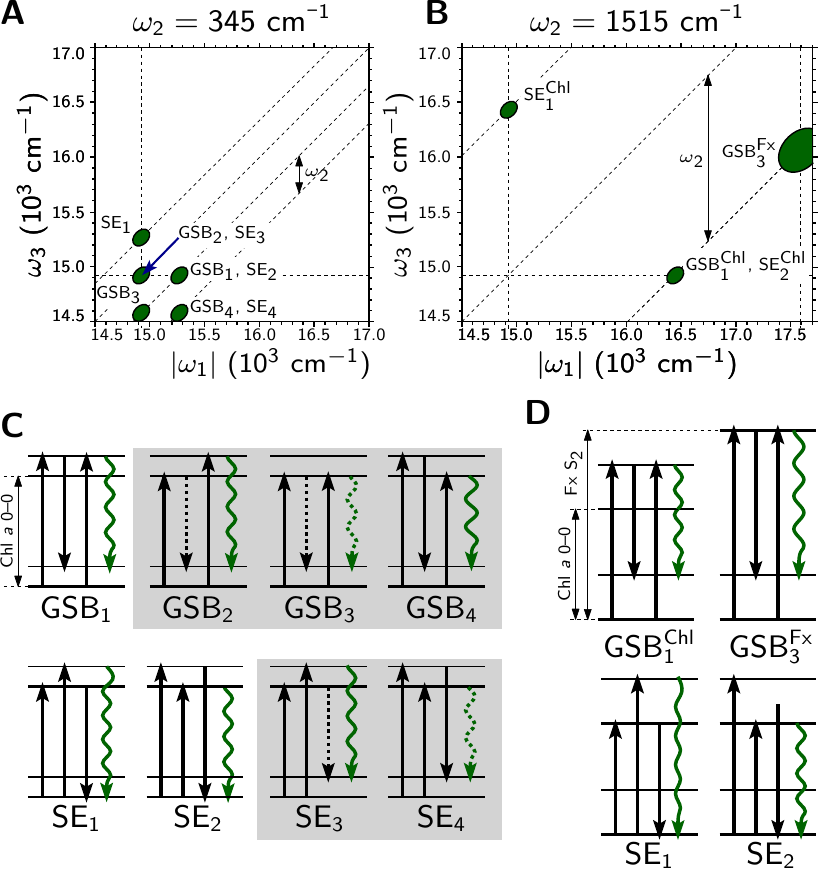}

\protect\caption{\label{fig:oscillations-scheme}Patterns of features in the Fourier
amplitude maps at $\omega_{2}=345\,\protect\icm$ (A) and $\omega_{2}=1515\,\protect\icm$
(B). Excitation pathways, leading to coherences in these maps are
shown in (C) and (D), respectively. Horizontal thick lines in (C)
and (D) show ground and excited state energies of Chl~$a$ and Fx.
Thin horizontal lines indicate the corresponding vibrational states.
Vertical straight arrows represent interactions; vertical wavy lines
indicate emission. Transitions, suppressed by the laser spectrum are
shown by dashed lines and the pathways, becoming negligible for the
higher frequency vibrational modes, are shown in the shaded background.}
\end{figure}

Frequencies, reported in these studies as well as obtained by our
experiment, are summarized in Table~\ref{tab:Frequencies}. The comparison
gives almost perfect agreement not only for the obtained strongest
coherences at around $\omega_{2}=255\,\icm$, $345\,\icm$, and $745\,\icm$
(deviations smaller than $11\,\icm$) but also for the weaker ones
with the largest deviation of $20\,\icm$ for the $\omega_{2}=1305\,\icm$
vibrational mode. The Fourier amplitude maps at $\omega_{2}=255\,\icm$,
$345\,\icm$, and $745\,\icm$ are shown in Fig.~\ref{fig:MAPS}.
All three of them clearly indicate the presence of vibrational and
vibronic coherences as the patterns of the oscillatory features can
be easily decomposed to 4 stimulated emission (SE) and 4 ground state
bleaching (GSB) contributions of a simple two-level system with one
vibrational state in the excited and ground states (see the scheme
and the corresponding excitation pathways in Fig.~\ref{fig:oscillations-scheme}A
and C). From the scheme it follows that weak features above the diagonal,
separated by exactly $\omega_{2}$ value, indicate vibronic excited
state coherences (${\rm SE_{1}}$ pathway).

\begin{table}
{\small{}}%
\begin{tabular}{ccccccccc}
\cline{1-4} \cline{6-9} 
\multirow{2}{*}{{\small{}FCP}} & \multicolumn{2}{c}{{\small{}Chl $a$}} & {\small{}Fx } & $\quad$ & \multirow{2}{*}{{\small{}FCP}} & \multicolumn{2}{c}{{\small{}Chl $a$}} & {\small{}Fx}\tabularnewline
\cline{2-3} \cline{7-8} 
 & HB & PP & RR &  &  & HB & PP & RR\tabularnewline
\cline{1-4} \cline{6-9} 
{\small{}135} & -- & -- & -- &  & -- & {\small{}864} & -- & --\tabularnewline
{\small{}190} & -- & {\small{}214} & -- &  & -- & {\small{}874} & -- & --\tabularnewline
\textbf{\small{}255} & {\small{}262} & \textbf{\small{}259} & -- &  & -- & {\small{}896} & -- & --\tabularnewline
-- & {\small{}283} & {\small{}300} & -- &  & {\small{}915} & \textbf{\small{}932} & {\small{}915} & --\tabularnewline
\textbf{\small{}345} & -- & \textbf{\small{}346} & -- &  & {\small{}985} & \textbf{\small{}994} & \textbf{\small{}982} & {\small{}1002}\tabularnewline
-- & {\small{}390} & -- & -- &  & {\small{}1025} & {\small{}1009} & {\small{}1043} & {\small{}1019}\tabularnewline
{\small{}410} & {\small{}425} & {\small{}407} & -- &  & {\small{}1060} & {\small{}1075} & {\small{}1084} & --\tabularnewline
{\small{}450} & {\small{}469} & -- & -- &  & -- & {\small{}1114} & {\small{}1124} & --\tabularnewline
{\small{}505} & {\small{}501} & -- & -- &  & \textbf{\small{}1155} & -- & -- & \textbf{\small{}1160}\tabularnewline
{\small{}530} & {\small{}521} & {\small{}519} & -- &  & -- & {\small{}1178} & {\small{}1175} & {\small{}1185}\tabularnewline
{\small{}555} & {\small{}541} & \textbf{\small{}565} & -- &  & {\small{}1215} & {\small{}1203} & -- & {\small{}1205}\tabularnewline
-- & \textbf{\small{}574} & -- & -- &  & {\small{}1255} & \textbf{\small{}1259} & {\small{}1252} & --\tabularnewline
{\small{}585} & {\small{}588} & -- & -- &  & {\small{}1305} & {\small{}1285} & -- & --\tabularnewline
-- & {\small{}607} & -- & -- &  & {\small{}1350} & {\small{}1340} & {\small{}1353} & --\tabularnewline
{\small{}630} & {\small{}638} & {\small{}621} & -- &  & {\small{}1375} & \textbf{\small{}1364} & -- & --\tabularnewline
{\small{}660} & -- & {\small{}667} & -- &  & -- & {\small{}1390} & -- & --\tabularnewline
-- & {\small{}692} & -- & -- &  & {\small{}1405} & {\small{}1411} & {\small{}1415} & --\tabularnewline
-- & {\small{}714} & -- & -- &  & -- & {\small{}1433} & -- & --\tabularnewline
\textbf{\small{}745} & \textbf{\small{}746} & \textbf{\small{}744} & -- &  & -- & {\small{}1455} & -- & --\tabularnewline
-- & {\small{}771} & -- & -- &  & {\small{}1460} & {\small{}1465} & {\small{}1460} & --\tabularnewline
-- & {\small{}791} & -- & -- &  & -- & {\small{}1504} & -- & --\tabularnewline
{\small{}810} & {\small{}805} & {\small{}799} & -- &  & \textbf{\small{}1515} & \textbf{\small{}1524} & {\small{}1516} & --\tabularnewline
-- & {\small{}819} & -- & -- &  & {\small{}1540} & -- & -- & \textbf{\small{}1536}\tabularnewline
{\small{}865} & {\small{}855} & -- & -- &  &  &  &  & \tabularnewline
\cline{1-4} \cline{6-9} 
\end{tabular}{\small \par}

\protect\caption{\label{tab:Frequencies}Summary of coherent oscillations ($\protect\icm$)
in the measured FCP 2D spectrum (first column). Vibrational frequencies
of chlorophyll $a$ obtained from the hole burning\citep{Gillie1989}
(HB, second column) and pump--probe\citep{Du2011} (PP, third row)
experiments. Vibrational frequencies of fucoxanthin from the resonant
Raman (RR) study\citep{FCP_RR_bba_2010} are shown in the fourth row.
Numbers in bold denote the five most intensive oscillations in 2D
and pump--probe experiments, two most intensive Raman modes of fucoxanthin
and frequencies from the hole-burning experiment, having Franck--Condon
factors higher than 0.02.}
\end{table}

The map at $\omega_{2}=745\,\icm$ lacks amplitude of oscillations
on the diagonal compared to the $\omega_{2}=345\,\icm$ or $\omega_{2}=255\,\icm$
maps due to the narrow bandwidth of the laser spectrum, suppressing
some specific coherences. That is, since Chl~$a$ 0--0 transition
is on the low-energy shoulder of the laser excitation spectrum, all
excitation pathways containing lower energy transitions (indicated
by the dashed arrows in Fig.~\ref{fig:oscillations-scheme}C) will
be substantially suppressed. These particular pathways in Fig.~\ref{fig:oscillations-scheme}C
are shown in a shaded background. 

For oscillations of higher frequencies, the mentioned contributions
will be completely excluded. Therefore, the dependency of amplitude
of oscillations in Fig.~\ref{fig:frobenius}A should be considered
carefully, as the peak heights there do not perfectly represent the
amplitudes of all summed pathways, but are rather suppressed in middle
and high frequencies.

The Fourier amplitude maps at $\omega_{2}=1155\,\icm$ and $1515\,\icm$
are shown in Fig.~\ref{fig:MAPS}. They both contain broad features
in the region of excitation frequency $\omega_{1}=16500-17000\,\icm$
and emission frequency $\omega_{3}\approx\omega_{1}-1155\,\icm$ and
$\omega_{3}\approx\omega_{1}-1515\,\icm$, respectively. With our
laser spectrum we could in principle excite the ${\rm S_{2}}$ states
of some of the ``red'' fucoxanthins and, possibly, get signatures
of carotenoid ground state coherences (from the ${\rm GSB_{3}^{Fx}}$
pathway shown in Fig.~\ref{fig:oscillations-scheme}D). These contributions
would appear as oscillations at the emission frequency, shifted down
from the diagonal by the frequency of the vibrational mode. Therefore,
we assign these features observed at $\omega_{2}=1155\,\icm$ and
$\omega_{2}=1515\,\icm$ to the ground state vibrational modes of
Fx at $1160\,\icm$ and $1536\,\icm$, respectively\citep{FCP_RR_bba_2010}. 

Apart from the Fx vibrational coherence, the Fourier amplitude map
at $\omega_{2}=1515\,\icm$ shows oscillations at $\omega_{3}=14935\,\icm$
and $\omega_{1}=16550\,\icm$ and a corresponding diagonal-symmetric
feature. They can be assigned to the Chl~$a$ vibrational and vibronic
coherences, respectively. Thus, the full pattern of features in the
map at $\omega_{2}=1515\,\icm$ can be explained by the scheme shown
in Fig\@.~\ref{fig:oscillations-scheme}B: the two diagonal-symmetric
narrow features come from the excited state vibronic and ground state
vibrational coherences of Chl~$a$ and the broad feature below the
diagonal is due to vibrational ground state coherences of Fx. The
other weak features at the emission frequency $\omega_{3}\approx16000\,\icm$
cannot be explained by any sort of excitation pathways; we assign
them to experimental artifacts coming from the nonrephasing contribution
(not shown).

The Fourier map at $\omega_{2}=1155\,\icm$ clearly lacks similar
features stemming from Chl~$a$ vibrational coherences and there
is no vibrational mode of similar frequency suggested by the hole-burning
or pump--probe experiments considered here\citep{Gillie1989,Du2011}.
However, vibrational Chl~$a$ mode of $1145\,\icm$ has been reported
in the fluorescence narrowing and resonant Raman study by Telfer~et~al\citep{Telfer2010}.
It is possible that we were not able to resolve signatures of this
particular vibrational coherence because of the suppression of high
frequency coherences by laser spectrum as discussed above. Alternatively,
there could be additional mechanisms, responsible for some vibrational
Chl~$a$ frequencies not being detected in FCP, e.g. vibronic coupling
between Chl~$a$ in FCP, adding more complexity to spectra, similar
to the enhancement of the ground state coherence proposed by Chenu~et~al\citep{Chenu2013},
or the anharmonicity of the electron-vibrational coupling recently
proposed for the LH2 antenae\citep{Rancova2014}. Obviously, the situation
calls for more delaited study of vibrational/vibronic coherences of
chlorophylls in coupled pigment--protein complexes.

After performing analysis of the Fourier maps of the other coherences,
we were not able to extract any oscillatory features that could be
assigned to the vibrational/vibronic coherences of Chl~$c$. This
is probably due to too weak signal. Indeed, the Chl~$c$ absorption
signal is at least 5 times weaker than the main absorption band of
Chl~$a$. Thus, the signal of coherent oscillations might become
too weak to be distinguished from the experimental noise. A close
inspection of oscillations around the proposed $\Qy$ transition band
of Chl $c$ ($\sim15600\,\icm$) revealed only very weak signatures
of vibrational coherences of Chl $a$ and Fx. The former might be
coming from the nonrephasing vibrational contribution and the latter
-- from the overlap of a wide peak of the fucoxanthin ground state
vibrational coherence. 

Obviously, the strongest high-frequency electronic coherence in the
spectra of FCP would be the coherence of $\Qy$ transitions of excitonically
coupled Chl~$c$ and Chl~$a$. From the position of the corresponding
peaks in the absorption spectrum, the frequency of such a coherence
should be around $715\,\icm$ or $915\,\icm$. By comparing the peak
heights and assuming that the Franck--Condon factors of vibrational
transitions are as small as $0.05$, it could easily be deduced that
the amplitude of electronic coherence should be higher than the vibrational
coherence of Chl~$c$. But the dephasing rate of the electronic coherences
is known to be much shorter than the vibrational ones\citep{Butkus_vibrations_theory_jcp_2014}.
This would prevent the electronic coherence to be captured by the
Fourier maps, except of a few cases. One of them is the inhomogeneous
distribution induced picosecond electronic coherence lifetime, as
it was recently shown by Dong et al.\citep{Dong2014}. In the inhomogeneously
broadened spectrum the sub-ensemble of molecules, having similar energies
provides long-lived coherent signal in the two-color photon echo experiment.
Such mechanism could also be significant in the 2D electronic spectroscopy,
however, it requires the absorption spectra of Chl~$c$ and Chl~$a$
to overlap significantly.

Another mechanism allowing for long-lived electronic coherences is
substantial electronic--vibrational mixing in the excited state, resulting
in a vibrational coherence lifetime borrowing mechanism\citep{Butkus2013}.
For this effect to be noticeable, the excitonic energy gap between
the excitonic states should closely match some strong discrete vibrational
mode. In our case, the proposed excitonic energy gap is close to a
very strong vibrational modes of $745\,\icm$ or $915\,\icm$ and
in the Fourier amplitude maps the ground state vibrational contribution
dominates. Therefore, if there are observable signatures of electronic
coherence beatings, they are hindered by strong signals of vibrational
coherences.

\subsection{Ultrafast Chl $c$--Chl $a$ transfer}

In our previous study of the 2D spectra of the FCP complex at room
temperature, the ultrafast energy transfer (occurring on timescale
of 60~fs) from excitonically coupled Chl~$c$ to Chl~$a$ was proposed\citep{Songaila_FCP_2013}.
It was obtained by tracing the time dynamics of the spectral feature
at $\omega_{3}=\omega_{1}=15680\,\icm$, assigned to the $\Qy$ transition
of Chl~$c$. Here we report that there are two peaks at $15610\,\icm$
and $15810\,\icm$ instead of one. This doublet was not resolved in
the experiment at the room temperature. Moreover, we found significant
differences in their decay dynamics. Considering times up to 30 ps,
the lower peak decays with the timescale of $320$ femtoseconds, while
the the shortest temporal dynamics of the upper peak is $3.9$~ps.
The timescales were obtained from temporal traces, taken by averaging
over the area of a circle of 100~$\icm$ radius in spectra and omitting
the first 30~fs, thus limiting amount spectral diffusion and pulse
overlap contributions appearing in peak dynamics. We can thus conclude
that the obtained two different Chl~$c$ states directly transfer
energy to Chl~$a$ on different timescales. Notice that energy transfer
via intermediate states (presumably from fucoxanthins) is not consistent
with fluorescence--excitation spectrum of FCP\citep{Papagiannakis2005},
which implies 100\% efficiency of Chl~$c$ to Chl~$a$ energy transfer,
and the fact that significant portions of excitation from Fx $\mathrm{S_{1}/ICT}$
(internal charge transfer) state, which could be suggested to be a
bridge between Chl~$c$ and Chl~$a$, are lost due to the decay
to the ground state\citep{Gelzinis_FCP_2D2color_BBA_2015}. 

Recent studies suggest that there are two Chl~$c$ molecules in the
complex\citep{FCP_RR_bba_2010}. Also, some of them can be of different
types (i.e. Chl~$c_{1}$, $c_{2}$ or $c_{3}$) that have slightly
different absorption spectra\citep{Kosumi_FCP_jpcl_2013}. Even if
the discussed spectral features are caused by the chlorophylls of
different types (thus, causing signals in different frequencies),
the temporal dynamics of the corresponding peaks indicate that their
interaction with the rest of the system or a surrounding environment
is different. The simplest picture is that one Chl~$c$ is situated
in the periphery of the FCP complex and is weakly coupled to the rest
of Chl~$a$, thus, the excitation transfer to the core pigments is
slow and we observe a picosecond decay of the peak at $15810\,\icm$.
At the same time, the protein environment causes it to have higher
energy. Another Chl~$c$, absorbing at $15610\,\icm$, must be coupled
to Chl~$a$ stronger for faster excitation transfer. Since the cross-peak
between these two transitions as well as the relevant excited state
absorption contribution is absent, chlorophylls $c$ in FCP are not
coupled. This implies that they probably are spatially well separated.

However, the ultrafast energy transfer from the lower energy Chl~$c$
state still remains an issue to be explained. The strength of coupling
between Chl~$c$ and Chl~$a$ is not known, although it may be estimated
from LHCII, which shares considerable sequence homology\citep{FCP_homology_1998}.
Couplings between chlorophylls in the LHCII complex were calculated
to be less than 100~$\icm$\citep{Duffy_etal_LHCII_couplings_jpcb_2013}.
Considering this and the fact that the transition dipole moment of
the $\Qy$ band is smaller for Chl~$c$ than for Chl~$a$, one can
be certain that coupling between Chl~$c$ and Chl~$a$ in FCP should
be even weaker. This is also supported by the absence of clear excitonic
cross-peaks between Chl~$a$ and Chl~$c$ states in our 2D spectra.

The F\"orster energy transfer could apply in this range of parameters.
However, it leads to slower picosecond transfer rates. The 320~fs
transfer timescale observed in 2D~ES can be understood by including
two ingredients in the consideration. First, the lower energy Chl~$c$
can be coupled to several Chl~$a$ molecules, thus, it transfers
energy to a few Chl~$a$ molecules at once. The rates effectively
sum-up and lead to the fast decay of the Chl~$c$ population. Second,
vibrations-assisted energy transfer enhancement might be significant
for the Chl~$c$--Chl~$a$ energy transfer. The energy gap of $715\,\icm$
is close to the strongest vibrational Chl~$a$ mode of $745\,\icm$,
thus allowing for electronic--vibrational resonance\citep{Plenio_noise_assisted_transport_jcp_2009,Womick2011,Kolli2012}.
Recently, specific vibrational modes were suggested to speed up the
initial charge transfer step in the photosystem II (PS II) reaction
center\citep{Fuller_NChem_2014,Romero2014}. Finally, this does not
contradict our previous suggestion that Chl~$c$ is coherently coupled
to Chl~$a$\citep{Songaila_FCP_2013}, however, the coherent mechanism
may be far more complicated and calls for future studies.

\begin{figure}
\includegraphics{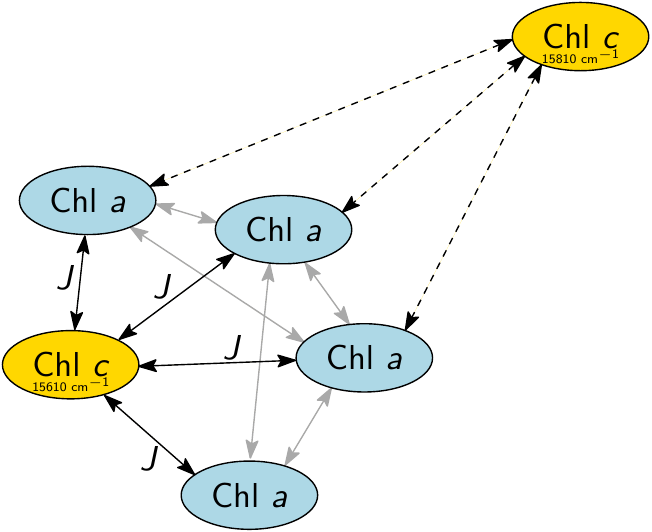}

\protect\caption{\label{fig:Schematic-illustration-of}Proposed functional scheme of
spatial arrangement of chlorophyll $c$ and $a$ molecules in the
FCP complex. Couplings between pigments are shown by solid lines.
Dashed lines indicate weaker coupling between spatially separated
Chl~$c$ absorbing at $15810\,\protect\icm$. Chl~$a$ molecules
shown here do not represent their actual number in FCP.}
\end{figure}

We propose a schematic spatial arrangement of Chl~$a$ and Chl~$c$
molecules (shown in Fig.~\ref{fig:Schematic-illustration-of}). This
functional scheme is comparable to the CP29 photosynthetic antenna
of photosystem II\citep{Gradinaru1998,Pan2011}. The structure of
the CP29 complex is similar to the light harvesting complex LHCII,
which shares the sequence homology with FCP. In CP29 there are eight
Chl $a$ and four Chl $b$ (the absorption maximum of Chl~$b$ is
around $15674\,\icm$, similarly as to Chl~$c$) and one more chlorophyll
binding site can be occupied by either type of chlorophyll\citep{Pan2011},
although previously it was assumed that there are 8 Chl~$a$ and
2 Chl~$b$ molecules\citep{Gradinaru1998}. However, in both models
one chlorophyll $b$ molecule is well spatially separated from the
other which is strongly coupled to the rest of chlorophylls $a$ of
the complex. Consequentially, two absorption bands at $15650\,\icm$
and $15385\,\icm$ are observed in the absorption and in the pump--probe
spectrum have completely different temporal dynamics: the energy transfer
from Chl~$b$ at higher energy is transferred to Chl~$a$ with the
rate of 350~fs, while transfer from the lower energy Chl~$b$ band
occurs in 2.2~ps\citep{Gradinaru1998}. This implies that having
two accessory chlorophylls that are in different spatial situations
might be a property shared by a number of light-harvesting complexes,
thus probably it has a functional significance.

In a recent study of FCP based on resonance Raman technique\citep{FCP_RR_bba_2010},
signatures of two distinct Chl~$c$ molecules were observed. In that
study a possible pigment organization scheme in FCP was discussed,
based on the pigment binding sites of the LHCII from plants\citep{Liu-Chang2004}.
It was suggested that one of the Chl~$c$ should be bound at either
the site of a614 (which is preserved in FCP) or a613 (not preserved
in FCP), while the other Chl~$c$ should be bound at the site of
b609 (preserved) or a604 (not preserved). For the faster transferring
Chl~$c$ the site of b609 would be a logical choice, as it is in
the vicinity of several pigments. Meanwhile, the higher energy (15810~$\icm$)
Chl~$c$ could indeed be situated at the site of a614, as that site
is relatively isolated and would be in accord with slower energy transfer
from this pigment. Since this pigment is situated at the periphery
of the complex, it might be responsible for the inter-complex Chl~$c$
to Chl~$a$ transfer in the FCP--PSII supercomplexes\citep{Akimoto_etal_FCP_bba_2014}.

\section{Conclusions}

In this work we studied low-temperature and high-resolution experimental
two-dimensional spectra of FCP complex. Short time (femtosecond) dynamics
of the spectra show complex oscillation patterns. They are dominated
by vibrational and vibronic coherences of Chl~$a$ and fucoxanthin;
electronic coherences could not be identified due to their short lifetime
and being hindered under high-amplitude oscillations originating from
Chl~$a$. Most notably, we have resolved two distinct states, that
can be assigned to chlorophylls $c$. These states were not found
in the previous studies of FCP and their energies were not known.
The two Chl~$c$ molecules transfer energy to Chl~$a$ with rates
that differ by about an order of magnitude. We suggest that the lower
energy Chl~$c$ molecule is situated in the core of the complex and
is coupled to an aggregated pool of Chl~$a$ molecules and the Chl~$c$-to-$a$
energy transfer might be vibronically-enhanced. The higher energy
Chl~$c$ should be in the periphery of the complex to transfer energy
on the picosecond timescale.

\section*{Acknowledgments}

Authors would like to thank Kerstin Pieper for preparing the FCP samples
and Egidijus Songaila for his work during the sample preparation procedure
for the 2D ES experiment.

\textcolor{black}{The research was partially funded by the European
Social Fund under the Global Grant measure. Authors acknowledge support
by LASERLAB--EUROPE project (grant agreement n\textdegree{} 228334,
EC's Seventh Framework Programme). }Work in Lund was also supported
by the Swedish Research Council and Knut and Alice Wallenberg Foundation.
V.~B. acknowledges support by project ``Promotion of Student Scientific
Activities'' (VP1-3.1-ŠMM-01-V-02-003) from the Research Council
of Lithuania. C.~B. acknowledges funding by the Deutsche Forschungsgemeinschaft
(Bu 812/4-1, 5-1). C.~B., A.~Gall and B.~R. acknowledges funding
from the EU (HARVEST Marie Curie Research Training Network (PITN-GA-2009-238017).
B.~R. acknowledges support from the European Research Council (ERC)
through an Advanced Grant, contract no. ERC-2010-AdG PHOTPROT.

\end{document}